

%
\def\section#1{\par \ifnum\lastpenalty=30000\else
   \penalty-200\vskip\sectionskip \spacecheck\sectionminspace\fi
   \gl@bal\advance\sectionnumber by 1
   {\pr@tect
   \xdef\sectionlabel{\ifcn@@ \chapterlabel.\fi
       \the\sectionstyle{\the\sectionnumber}}%
   \wlog{\string\section\space \sectionlabel}}%
   \noindent {\caps\enspace\sectionlabel.~~#1}\par
   \nobreak\vskip\headskip \penalty 30000 }
\def\newpart#1{\par \ifnum\the\lastpenalty=30000\else
   \penalty-200\vskip\sectionskip \spacecheck\sectionminspace\fi
   \wlog{\string\section\ \chapterlabel \the\sectionnumber}
   \global\advance\sectionnumber by 1  \noindent
   {\bf #1}\par
   \nobreak\vskip\headskip \penalty 30000 }
\let\eqnameold=\eqname
\def\draft{\def\eqname##1{\eqnameold##1:{\tt\string##1}}
              \let\eqnalign = \eqname}

\def\chapdown{\hbox to \hsize{\null}   \vskip 0.25in}

\def\Buildrel#1\under#2{\mathrel{\mathop{#2}\limits_{#1}}}
\def\a{\alpha}

\def\s{\sigma}

\def\rarr{\rightarrow}

\def\half{{1\over 2}}

\def\bar#1{\overline{#1}}
\def\bold#1{\setbox0=\hbox{$#1$}%
     \kern-.025em\copy0\kern-\wd0
     \kern.05em\copy0\kern-\wd0
     \kern-.025em\raise.0433em\box0 }

\def\dslash{\not{\hbox{\kern-2pt $\partial$}}}
\def\Dslash{\not{\hbox{\kern-4pt $D$}}}
\def\Qslash{\not{\hbox{\kern-4pt $Q$}}}
\def\pslash{\not{\hbox{\kern-2.3pt $p$}}}
\def\kslash{\not{\hbox{\kern-2.3pt $k$}}}
\def\qslash{\not{\hbox{\kern-2.3pt $q$}}}

\def\refmark#1{\attach{\scriptstyle [ #1 ] }}
\def\etal{{\it et~al.}}

%



%
\pubnum{6228}
\pubtype{(T/E)}
\titlepage
\title{Minijet Corrections to Higgs Production}

\author{Vittorio Del Duca and Carl R. Schmidt\doeack}
\SLAC
\abstract{We study higher order corrections to Higgs production
with an associated jet at SSC energies,
using the resummation of the leading
logarithmic contributions to multiple gluon emissions due to Lipatov
and collaborators. We find a considerable enhancement of Higgs production
at large transverse momenta.}
\submit{{\sl Physical Review} {D}}
\vfil
\centerline{PACS numbers: 14.80.Gt, 13.87.Ce, 12.40.Gg, 13.85.Qk}

\endpage
\chapter{Introduction}
One of the most important open questions in the Standard Model is the
cause of the electroweak symmetry breaking.  Both the SSC and the LHC
are being built with the resolution of this puzzle in mind.  The
simplest explanation of the symmetry breaking is by a fundamental
$SU(2)$ doublet scalar field, which leaves a neutral scalar, the Higgs
boson, as a remnant of the symmetry breaking.  This simple theory is
the experimental standard against which any model of symmetry breaking
will be measured.  Thus, it is important that we completely understand
the production environment of the Higgs boson at these hadron
colliders.

At the SSC and the LHC, the Higgs boson is predominantly produced by
gluon fusion.  At lowest order $\alpha_s^2$, it is produced with zero
transverse momentum.  The complete order $\alpha_s^3$ corrections have
been calculated\Ref\Zerwas{S. Dawson, \sl Nucl. Phys. \bf B359, \rm
(1991) 283;\hfill\break
A. Djouadi, M. Spira, and P.M. Zerwas, \sl Phys. Lett. \bf B264, \rm
(1991) 264;\hfill\break
D. Graudenz, M. Spira, and P.M. Zerwas, \sl Phys. Rev. Lett. \bf 70,
\rm (1993) 1372.} with an increase in the
total cross section by a factor of 1.5 to 1.7.  In addition, production
of Higgs bosons with nonzero transverse momentum have been studied at
order $\alpha_s^3$.  Higgs bosons produced with sizable $p_\perp$ may
be phenomenologically important both in the intermediate mass region,
where rare decay modes such as $\gamma\gamma$ and $\tau^+\tau^-$ must
be used to observe the Higgs\rlap,\Ref\EHSV{R.K. Ellis, I. Hinchliffe, M.
Soldate, J.J. Van Der Bij, {\sl Nucl. Phys.} {\bf B297}, 221 (1988).}
and for heavier Higgs bosons where the vector boson decay channels may be
used\rlap.\Ref\BG{U. Baur and E.W.N. Glover, {\sl Nucl. Phys.} {\bf B339}, 38
(1990).}\  At small transverse momentum it is
necessary to resum large logarithms in $m_H^2/p_\perp^2$ in order to
obtain a physical $p_\perp$
distribution\rlap.\REFS\HN{I. Hinchliffe and S.F. Novaes, \sl Phys. Rev.
\bf D38, \rm (1988) 3475.}
\REFSCON\kauf{R.P. Kauffman, \sl Phys. Rev. \bf D44, \rm (1991) 1415;
\bf D45, \rm (1992) 1512.}
\REFSCON\yuan{C.P. Yuan, \sl Phys. Lett. \bf B283, \rm (1992) 395.}
\refsend

\REF\CE{J.C. Collins
and R.K. Ellis, \sl Nucl. Phys. \bf B360 \rm (1991) 3; \hfill\break
S. Catani, M. Ciafaloni and F. Hautmann, \sl Nucl. Phys. \bf B366 \rm (1991)
135.}
\REF\MN{A.H. Mueller and
H. Navelet, \sl Nucl. Phys. \bf B282 \rm (1987), 727.}
\REF\BFKL{L.N.
Lipatov, \sl Sov. J. Nucl. Phys. \bf 23 \rm (1976) 338; \hfill\break
E.A. Kuraev, L.N. Lipatov and V.S. Fadin, \sl Sov. Phys. JETP \bf
44 \rm (1976) 443, \bf 45 \rm (1977) 199; \hfill\break
Ya. Ya. Balitskii and L.N. Lipatov, \sl Sov. J. Nucl. Phys. \bf 28
\rm (1978) 822.}
At the large center of mass energy $\sqrt{s}$ of the SSC and the LHC
there are also important contributions from events with multiple
final-state partons.  These arise in a new kinematical region, the
semihard region, characterized by scattering processes with
$\sqrt{s}$ much larger than the momentum transfer $Q$, $s >> Q^2
>> \Lambda_{QCD}^2$. The study of this region is
theoretically challenging because, in the perturbative expansion of
cross sections, there appear coefficients containing logarithms of
large ratios of kinematical
invariants, of the order of the rapidity interval in the scattering
process.  In the inclusive production of the Higgs boson,
the large ratios of kinematical invariants appear in the
small-$x$ evolution of the structure functions, thus requiring a more
sophisticated analysis than the usual DGLAP evolution\rlap.\refmark
\CE

In the context of jet production, Mueller and Navelet\refmark\MN\ were
able to disentangle the small-$x$ evolution of the structure
functions from the appearance of large ratios of kinematical invariants
in the parton subprocess by proposing to tag
two jets at the extremes of the Lego plot in azimuthal angle and
rapidity, with large transverse momenta $p_{\perp}$, and longitudinal
momentum fractions $x_A$, $x_B$ large enough that the parton structure
functions could be computed from the ordinary DGLAP evolution. Thus,
fixing the parton center of mass energy $\hat s = x_A x_B s$, the
semihard region is realized also at the parton level $\hat s >> Q^2$,
and the large logarithmic terms $\ln(\hat s/Q^2)$ appear in the parton
cross section. To deal with these, Mueller and Navelet used the
Balitsky-Fadin-Kuraev-Lipatov theory (BFKL)\rlap,\refmark\BFKL\
which systematically resums the leading powers in the rapidity interval
by using a multigluon amplitude where the rapidity interval between the
{\it tagging jets} is filled with gluons, the {\it minijets}, whose
spacing in rapidity is approximately uniform.

For the Higgs boson this program
can be implemented by considering the production of
the Higgs boson and a jet at the extremes of the Lego plot, with the
rapidity interval between them filled with minijets, and
the Higgs boson and the jet tagged at large transverse
momenta. Since in gluon fusion the Higgs boson is
produced via a quark loop, and the
fermion-Higgs coupling is proportional to the quark mass, the only
contribution to the quark loop which is numerically important is due to
the top quark. Thus, two more scales, the Higgs boson and the top quark
masses, enter the kinematics, and in order to avoid complications with
the evolution of the structure functions\refmark\CE\ and apply the BFKL
analysis, we must require that $\hat s>>p_\perp^2,m_H^2,m_t^2$.
However, when $p_{\perp}^2 << m_H^2$
doubly logarithmic terms $\ln^2(m_H^2/p_{\perp}^2)$ appear, which more
properly belong to the evolution of the structure functions
\refmark{\HN-\yuan}\ and may
void the BFKL analysis. So in order to single out the strong
rapidity-ordering regime characteristic of the BFKL analysis, the Higgs
and the jet transverse momenta must be comparable to the Higgs mass
$p_{\perp}^2 \simeq m_H^2$.

In section 2, we consider the inclusive production of the Higgs boson and
a tagging jet, $p\, p \rarr H + {\rm jet} + X$ in this kinematic
situation.
We recall the exact Born-level calculation of Refs.~\EHSV\ and \BG\
and consider its large-rapidity limit, and then we compute the
leading logarithmic corrections at large rapidity,
using the BFKL analysis. In section 3, we
present numerical results for the inclusive Higgs-jet production, with
or without integrating over the Higgs and jet transverse momenta, and
compare the Born-level to the minijet-corrected calculations. Our
conclusions are presented in section 4.

\chapter{The Higgs $+$ Jet Inclusive Cross section}
We are going to study the semi-inclusive process $pp\rarr H + {\rm jet}
+ X$ in the semihard regime defined by $\hat s>>Q^2$,
with $Q^2$ being a typical momentum scale in the event, $Q^2\approx m_H^2,
m_t^2, p_\perp^2$. The tagging jet is required to
ensure that we have an event with a large rapidity interval $y=y_J-y_H
\approx \ln(\hat s/Q^2)$.  Other relevant parameters in the event are
the Higgs and jet transverse momenta $p_{H\perp},p_{J\perp}$, their
relative azimuthal angle $\phi$ and the average rapidity
$\bar y=(y_J+y_H)/2$.

In the semihard, large-$y$ regime we can write the cross section:
$${d\s\over dp_{H\perp}^2 dp_{J\perp}^2 d\phi dy d\bar y}\
=\ \sum_{ij}x_1x_2\,f_i(x_1)f_j(x_2)\,{d\hat\sigma_{ij}\over
dp_{H\perp}^2 dp_{J\perp}^2 d\phi}\ .
\eqn\general$$
The parton subprocess cross section $d\hat\sigma_{ij}/
dp_{H\perp}^2 dp_{J\perp}^2 d\phi$ contains the sum over all additional
particles (i.e.~minijets) in the event.  The factorization of the
minijets into the subprocess cross section is possible,
because at large $y$ the initial parton momentum
fractions $x_1$ and $x_2$ are fixed in terms of the Higgs and jet
momenta, and are essentially independent of the particles filling the
rapidity interval.  We will arrive at this cross section in several
steps, starting with the exact Born level cross section, taking it to
the $y>>1$ limit, and finally filling in the rapidity interval
with the minijets.  Of course, the analysis can be applied equally well
for $y<<-1$, by reflecting along the beam axis.

{\it i})\ \undertext{Born Level Cross section}.
At the Born level the Higgs boson and the parton jet are produced
back-to-back.  The exact lowest order cross section can be put in the
form \general\ with the replacement
$${d\hat\sigma_{ij}\over dp_{H\perp}^2 dp_{J\perp}^2 d\phi}
\ \Rightarrow\ {d\hat\sigma_{ij}\over d\hat
t}\,\delta(p_{H\perp}^2-p_{J\perp}^2)\,\delta(\phi-\pi)\ .\eqn\twobody$$
The parton momentum fractions and the subprocess invariants at this
level are given by:
$$\eqalign{
x_1\ &=\ {e^{{\bar y}}\over\sqrt{s}}(p_\perp e^{y/2}
+m_{H\perp}e^{-y/2})\cr
x_2\ &=\ {e^{-\bar y}\over\sqrt{s}}(p_\perp e^{-y/2}
+m_{H\perp}e^{y/2})\cr
\hat s\ &=\ x_1x_2s\ =\ p_\perp^2 + m_{H\perp}^2 + 2p_\perp
m_{H\perp}\cosh(y)\cr
\hat t\ &=\ -p_\perp^2-p_\perp m_{H\perp}e^{-y}\cr
\hat u\ &=\ -p_\perp^2-p_\perp m_{H\perp}e^{y}\ ,}\eqn\invar$$
where $m_{H\perp}=(p_\perp^2+m_H^2)^{1/2}$ and $p_\perp=p_{H\perp}=
p_{J\perp}$.
We use the lowest order Standard Model calculation for the subprocess
cross sections $d\hat\sigma_{ij}/d \hat t$ in Ref.~\EHSV.

\FIG\feyn{Higgs + jet production amplitude in the large-$y$ limit at
(a) the Born level and (b) with minijet corrections.}
{\it ii})\ \undertext{Large-$y$ Born Cross section}.
We now investigate the lowest order cross section when the rapidity interval
$y$ is large.    For $y>>1$ the lowest order amplitude is
dominated by diagrams with gluon-exchange in the $t$-channel as in
Fig.~\feyn(a).
In this limit the only subprocesses that contribute are $gg\rarr gH$
and $q(\bar q)g\rarr q(\bar q)H$.  We obtain
$${d\hat\sigma_{gg}\over d\hat t}\ =\ C\,{|{\cal F}
(p_\perp^2)|^2\over p_\perp^2}\ ,\eqn\largey$$
where
$$C\ =\ {N_c\a_s^3\a_W\over(N_c^2-1)128\pi M_W^2}\ .\eqn\const$$
Similarly, we find
$$ {d\hat\sigma_{qg} \over d\hat t}\ =\ {d\hat\sigma_{\bar qg} \over d\hat
t}\ =\ {C_F \over C_A} \, {d\hat\sigma_{gg} \over d\hat t}\ , \eqn\effec$$
with $C_F/C_A = (N_c^2-1)/2N_c^2 = 4/9$ the ratio of the Casimir
operators. The form factor
${\cal F}(p_\perp^2)$ is given in the appendix for both a scalar and a
pseudoscalar Higgs boson. We have checked that \largey\ agrees with
previous results\refmark\kauf\ if we take the
additional limit $p_\perp^2<<m_H^2,m_t^2$.
The parton momentum fractions in the large-$y$ limit are
$$\eqalign{
x_1\ &=\ {p_{J\perp}\over\sqrt{s}}e^{(\bar y+y/2)}\ =\
{p_{J\perp}\over\sqrt{s}}e^{y_J}\cr
x_2\ &=\ {m_{H\perp}\over\sqrt{s}}e^{(-\bar y+y/2)}\ =\
{m_{H\perp}\over\sqrt{s}}e^{-y_H}\ .}\eqn\largeyx$$
Equation \largeyx\ is also valid in the large-$y$ limit when higher-order
corrections are included, so that $p_{H\perp} \neq p_{J\perp}$.

{\it iii})\ \undertext{Minijet-corrected Cross section}.
As discussed in the introduction, going to higher
orders in the coupling constant, i.e.~to multiple parton emission,
we encounter large logarithmic contributions. In the semihard regime,
the BFKL theory\refmark\BFKL\
systematically resums the leading logarithmic terms $\ln(\hat{s}/p_{\perp}^2)$
by using a multigluon amplitude where the rapidity interval between the
Higgs boson and the tagging jet is filled with gluons, strongly ordered
in rapidity. This amplitude is shown in Fig.~\feyn(b), where the thick
line represents the resummation of the virtual radiative corrections,
whose effect is to reggeize the gluons exchanged in the $t$ channel.
The real gluons are inserted on these using the Lipatov effective
3-gluon vertex\rlap.\refmark\BFKL\ \ The BFKL multigluon amplitude is then
put in a rapidity-ordered phase space, the rapidities of the
gluons are integrated out, and the dependence of the cross section on
the gluon transverse momenta is reduced to the resolution of an
integral equation. Its solution is then convoluted with the Higgs-boson
production vertex on one side and the jet emission vertex on the other
side of the rapidity interval to give the minijet-corrected parton cross
section for producing the Higgs boson and a jet:
$$ {d\hat{\sigma}\over d^2p_{H\perp} d^2p_{J\perp}}\ =\ {2C\over\pi} \,
{|{\cal F}(p_{H\perp}^2)|^2 \over p_{J\perp}^2}
f(y,p_{H\perp},p_{J\perp})\ .
\eqn\cross$$
In this equation
$f(y,p_{H\perp},p_{J\perp})$ is the Laplace transform in the rapidity
interval $y$,
$$ f(y,p_{H\perp},p_{J\perp}) = \int {d\omega \over 2\pi i} e^{\omega y}
f_{\omega}(p_{H\perp},p_{J\perp}), \eqn\laplace$$
of the solution of the BFKL integral equation
$$ f_{\omega}(p_{H\perp},p_{J\perp}) = {1 \over (2\pi)^2}
\sum_{n=-\infty}^{\infty}
e^{in(\phi-\pi)} \int_{-\infty}^{\infty} d\nu {(p_{H\perp}^2)^{-1/2+i\nu}
(p_{J\perp}^2)^{-1/2-i\nu} \over \omega-\omega(n,\nu)}. \eqn\lip$$
The eigenvalue of the integral equation $\omega(n,\nu)$ is
$$ \omega(n,\nu) = {2 N_c \alpha_s \over\pi} \bigl[ \psi(1) - Re\,\psi
   ({|n|+1\over 2} +i\nu) \bigr], \eqn\eigen$$
with $\psi$ the logarithmic derivative of the Gamma function.
Substituting \laplace\ and \lip\ in \cross, and doing the integral over
$\omega$, the minijet-corrected parton cross section becomes
$$ {d\hat{\sigma}\over dp_{H\perp}^2 dp_{J\perp}^2 d\phi} = {C \over
2\pi^2} {|{\cal F}(p_{H\perp}^2)|^2 \over p_{H\perp} \, p_{J\perp}^3}
\sum_n e^{in(\phi-\pi)} \int_0^{\infty} d\nu e^{\omega(n,\nu)\, y}
\cos\left(\nu \, \ln{p_{H\perp}^2 \over p_{J\perp}^2} \right).
\eqn\mini$$
If we integrate over the azimuthal angle $\phi$ in \mini, only the
$n=0$ term survives.

{\it iv})\ \undertext{Minijet-corrected Cross section in the Saddle
Point Approximation}.\hfill\break
At very large values of the rapidity interval $y$, the correlations
between the Higgs boson and the jet are washed out by the random walk in
transverse momentum space of the gluons exchanged in the $t$ channel.
This can be seen most easily by evaluating \mini\ in the saddle-point
approximation. The contribution of \eigen\ to this equation is
dominated by $n = 0$ and is strongly peaked near $\nu = 0$.
Thus we keep only the first term in the Fourier expansion in $\phi$, and
expand $\omega(\nu) = \omega(0,\nu)$ about $\nu = 0$
$$ \omega(\nu) = A - B\nu^2 +\cdots, \eqn\ome$$
with
$$ A = {4 N_c \alpha_s \over \pi} \ln2, \quad B = {14 N_c \alpha_s
\over \pi} \zeta(3). \eqn\nuzero$$
Then we can evaluate \mini\ using the saddle-point approximation for
the integral over $\nu$, to obtain
$$ {d\hat{\sigma}\over dp_{H\perp}^2 dp_{J\perp}^2 d\phi} = {C\over 4\pi}
\, {e^{A\, y}\over \sqrt{B \, \pi \, y}} \,
\exp\left(-{\ln^2(p_{H\perp}^2/p_{J\perp}^2) \over 4 B y}\right) \,
{|{\cal F}(p_{H\perp}^2)|^2 \over p_{H\perp} \, p_{J\perp}^3}.
\eqn\asymp$$
The exponential growth of \asymp\ with the rapidity interval $y$ is due
to the production of the minijets.

\chapter{Numerical Results}

We now examine numerically the effects of the minijets on Standard Model
Higgs production  at SSC energies. Throughout this section we will
use a representative Higgs mass of 100 GeV and a top mass of 150 GeV,
and the SSC center of mass energy $\sqrt{s}$ = 40 TeV.  The qualitative
results do not depend strongly on these values of $m_H$ and $m_t$.
Except where indicated we set $\bar y = 0$ and
observe the cross sections as a function of the rapidity interval
$|y|$.  We obtain a factor of two by including both positive and
negative $y$.  We use the ``average'' parton density functions
of Diemoz \etal,\Ref\DFLM{M. Diemoz, F. Ferroni, E. Longo, G. Martinelli,
{\sl Z. Phys.} {\bf C39}, (1988) 21.} and we evaluate the QCD coupling
$\alpha_s$ and the structure functions at a scale $Q^2=m_H^2$.
For $p_\perp>50$ GeV and $|y|>4$ the structure functions are always
evaluated at $x>10^{-2}$, so we are justified in using the DGLAP
evolution in this region of phase space.

\FIG\xsec{Inclusive Higgs-jet production at the SSC, as a function of
the rapidity interval $y$. The dashed and dotdashed lines are
respectively the exact and large-$y$ Born cross sections, the solid
line is the minijet-corrected cross section, and the dotted line is the
saddle point approximation to the solid line. The kinematic parameters
are described in the text.}
\FIG\kfactor{$K$-factor, as a function of the rapidity interval $y$.
The solid line represents the ratio of the minijet-corrected cross
section to the large-$y$ Born cross section, and the dotted line is its
saddle point approximation.}
In Fig.~\xsec\ we present the inclusive Higgs-jet cross section,
integrated over the azimuthal angle $\phi$ and over both transverse
momenta with a cutoff of $p_{\perp\rm min}= 50$ GeV.  We present it in
the four approximations ({\it i-iv}) given in section 2.  From the plot
we see that the large-$y$ Born cross section is a good
approximation to the exact Born level cross section for $|y|
\gsim4$. The minijet-corrected cross section is
typically enhanced over these lowest order cross sections
by a factor of 2-3 for $|y|\gsim4$.  We define the $K$-factor as
the ratio of the
minijet-corrected cross section to the large-$y$ Born cross section
$$ K\  =\ {d\sigma ({\rm minijet})\over dy d\bar y} \bigg/\,\,
{d\sigma ({\rm large}\!-\!y)\over dy d\bar y}, \eqn\kfact$$
and plot it in Fig.~\kfactor.  $K$ is defined so that $K\rightarrow1$
as $y\rightarrow0$.  The minijet enhancement is often plotted on a
logarithmic scale\refmark\MN; on such a plot, the exact integral
approaches the saddle point approximation,
since the relative error tends to zero at large $\hat s$.

\FIG\ptfour{Higgs and jet $p_\perp$ distributions at (a) $|y|$ = 4 and
(b) $|y|$ = 8. The dashed and dotdashed lines
are respectively the $p_\perp$ distributions for the exact and the
large-$y$ Born cross section, for which $p_{H\perp} = p_{J\perp}$.
The solid lines are the jet and Higgs $p_\perp$
distributions for the minijet-corrected cross section. Notice that in
(b) the dashed and dotdashed lines completely overlap.}
We next show the $p_\perp$ distributions of the Higgs and the tagging
jet for rapidity intervals of $|y|=4$ and $|y|=8$ in Figs.~\ptfour(a)
and \ptfour(b), respectively.
The other momentum is integrated over $p_\perp>p_{\perp\rm
min}=50$ GeV.  Of course, for the Born cross section we have
$p_{H\perp}=p_{J\perp}$, so the transverse momentum distributions of
the Higgs and the jet are identical.  However, with the inclusion of
the minijets, this is no longer true.  In fact the $p_{H\perp}$
distribution is considerably flatter than either the lowest order
result or the minijet-corrected $p_{J\perp}$ distribution.

This effect can be understood by referring to Fig.~\feyn(b) and
analyzing the various terms in the minijet-corrected cross section
\cross.  When convoluted with the structure functions we find
$$\sigma\ \approx\ \biggl[x_2f(x_2)\Bigl|{{\cal F}(p_{H\perp}^2)\over v}
\Bigr|^2
\biggr]\,f(y,p_{H\perp},p_{J\perp})\,\biggl[x_1f(x_1)\Bigl({1\over
p_{J\perp}}\Bigr)^2\biggr]\ ,\eqn\vertex$$
where the first bracketed term is associated with the Higgs boson
production vertex and the last bracketed term is associated with the
jet emission vertex.  Note that the $1/p_{J\perp}^2$ comes from the jet
vertex, whereas the form factor ${\cal F}(p_{H\perp}^2)$ is relatively
constant in the region of interest.  For very small rapidities we
approach the Born cross section \largey\ with
$$f(y,p_{H\perp},p_{J\perp})\ \rightarrow\
\delta(p_{H\perp}^2-p_{J\perp}^2)\,\delta(\phi-\pi)\ ,\eqn\smally$$
so that both the Higgs boson and jet $p_\perp$ distributions fall as
$1/p_\perp^2$.  In addition, both $x_1$ and $x_2$ increase with
$p_\perp$ via \largeyx\ so that there is an additional suppression
from both parton density factors $(xf(x))^2$ as $p_\perp$
increases.  However, for very large rapidities we have
$$f(y,p_{H\perp},p_{J\perp})\ \rightarrow\
\sim (p_{H\perp}p_{J\perp})^{-1}\ ,\eqn\largey$$
and the Higgs boson and the tagging jet become uncorrelated.  Now
the Higgs boson distribution falls slower as $1/p_{H\perp}$,
while the jet distribution falls faster as $1/p_{J\perp}^3$.
In addition both distributions are now only suppressed by one factor of
$xf(x)$.  These effects combine to strongly broaden the Higgs
$p_\perp$ distribution. For the tagging jet the two
effects tend to cancel and its $p_\perp$ distribution is much less modified.

Comparing Figs. \ptfour(a) and (b), we can also observe some subtleties that
have been neglected in the arguments of the last paragraph.  For example,
the separation between the Higgs and jet $p_{\perp}$ distributions at
large $p_\perp$ shrinks as $|y|$ goes from
4 to 8.  This is due to the fact that both quarks and gluons can
enter the parton scattering on the jet side, while only gluons can
enter on the Higgs side. Thus, the suppression from the structure
functions is more severe on the Higgs side than on the
jet side as $x$ grows according to \largeyx.

Finally, we discuss the effects of varying the Higgs and top-quark
masses in the $p_{\perp}$ distributions of Fig. \ptfour.
Although the change in the contribution of the minijets is minor,
there will still be some overall effects on the distributions.  For
example, because we have chosen the
factorization and renormalization scales to be fixed at the Higgs mass,
as $m_H$ grows the coupling constant $\alpha_s$ and the structure
functions become smaller, while the parton momentum fraction $x$
increases.  Consequently, the absolute scale of the
$p_{\perp}$ distributions will decrease.
Changing the top-quark mass has a different effect.
The form factor ${\cal F}(p_{H\perp}^2)$ is roughly constant for
$p_{\perp} \lsim m_t$, but falls off as $m_t^2/p_{\perp}^2$ times
logarithms for $p_{\perp} \gsim m_t$. Thus, while the jet $p_{\perp}$
distribution becomes independent of $m_t$ at very large $y$,
the Higgs and Born $p_{\perp}$ distributions fall more steeply
for $p_{\perp} \gsim m_t$.  For a given fixed
$p_{\perp}$ larger than the top mass, the
Higgs and Born $p_\perp$ distributions will increase with $m_t$.

\FIG\ptcorr{Jet $p_\perp$ distribution at a fixed Higgs transverse
momentum of 125 GeV.
{}From top to bottom, the solid lines are the jet $p_\perp$
distributions for the minijet-corrected cross section at $|y|$ = 2, 4,
6 and 8. The dotted lines are the saddle point approximation to the
solid lines.}
The disappearence of correlations as $|y|$ increases
can be seen dramatically in Fig.~\ptcorr\
where we plot the transverse momentum distribution of the tagging jet at
a fixed Higgs transverse momentum of $p_{H\perp} = 125$ GeV.  For a
rapidity of $|y|=2$ the cross section is strongly peaked near
$p_{H\perp}=p_{J\perp}$.  As the rapidity is increased there is a
diffusion of the jet momentum away from the Higgs momentum until
the peak disappears completely for $|y|=8$.  Note that the saddle point
approximation removes the correlation for all values of $|y|$.
There is a similar diffusion of the azimuthal angle away from $\phi=\pi$
as the rapidity increases.

\FIG\yhzero{Higgs $p_\perp$ distribution at $y_H=0$. The dashed and
dotdashed lines are the distributions for the exact Born cross section
with, respectively, no cuts on the jet rapidity $|y_J|$, and $|y_J| > 4$.
The solid line is the distribution for the minijet-corrected cross
section with $|y_J| > 4$ and $p_{J\perp}>50$ GeV.}
In Fig.~\yhzero\ we consider the $p_{H\perp}$ distribution of
centrally produced Higgs bosons.  We fix $y_H=0$ while integrating over
the rapidity and transverse momentum of the tagging jet.  The dashed
line is the lowest order Born calculation with no cuts on the jet
rapidity, while the dotdash line has a cut of $|y_J|>4$.  The jet
transverse momentum is fixed at $p_{J\perp}=p_{H\perp}$ in these
Born-level curves.  The solid
line is the minijet-corrected cross section with cuts of $|y_J|>4$ and
$p_{J\perp}>50$ GeV.  We apply these cuts in order to be confident that
we are in a kinematical region where the minijet-corrected cross
section is a good approximation.  As in the previous plots we see that
the minijet-corrected $p_{H\perp}$ distribution is much broader than
the lowest order distribution.  At $p_{H\perp}=400$ GeV the minijet
cross section with the kinematic cuts is over two orders of magnitude
larger than the Born cross section with cuts.  Moreover, it is almost
within a factor of 2 of the lowest order Born cross section without any
cuts at all.

\chapter{Discussion and Conclusions}

In this paper we have calculated the contribution of minijets to Higgs
production in the semihard regime.  The major phenomenological
consequence of the minijets at SSC energies is to produce a substantial
enhancement of Higgs bosons at large transverse momenta.  Although our
results are strictly valid for the semi-inclusive process $pp\rarr H +
{\rm jet} + X$ at large $y$, we have seen that these events may still
make a sizable contribution to the inclusive process $pp\rarr H + X$
for large values of $p_{H\perp}$.  The enhancement occurs because in
the minijet-corrected cross section the falloff with $p_{H\perp}$ is
much slower than the falloff with $p_{J\perp}$. This suggests that
there may be a sizable contribution from events with a Higgs boson
produced at large transverse momentum, balanced by the collective
$p_\perp$ of several less energetic
jets.  These events should appear first at order $\alpha_s^4$ in the
$gg\rightarrow Hgg$ and $gq\rightarrow Hgq$ cross sections.  The former
has been calculated at the Born level in the large $m_t$
limit by Dawson and Kauffman\rlap,\Ref\DK{S. Dawson and R.P.
Kauffman, \sl Phys. Rev. Lett. \bf 68, \rm (1992) 2273.}\ who plotted
the cross section with a single $p_\perp$ cutoff on both the outgoing
gluons and the Higgs boson.  It would be interesting to study this
$gg\rightarrow Hgg$ cross section with the Higgs and gluon transverse
momenta varied independently in order to check for an enhancement of
events at large $p_{H\perp}$.

\noindent{\bf Acknowledgements}

We wish to thank bj. Bjorken, Al Mueller, Michael Peskin, Wai-Keung
Tang and Peter Zerwas for useful criticism and discussions.

\appendix

The analysis in this paper is valid both for a scalar Higgs boson and
for a pseudoscalar Higgs boson, as would occur in
two-Higgs-doublet models.  The only change in our results is in the
form factor ${\cal F}(p_\perp^2)$.  We present the analytic form of
${\cal F}(p_\perp^2)$ in both cases below.

{\it i)}\ \undertext{Scalar Higgs}.
The scalar Higgs coupling to the top quark is $-S(m_t/v)H\bar\psi\psi$,
where $S$ is a normalization constant which equals 1 for the Standard
Model Higgs.  The form factor is
$$\eqalign{
{\cal F}_s(p_\perp^2)\ =\ S\,\Bigl({4m_t^2\over m_{H\perp}^2}\Bigr)
\Biggl\{&-2\,-\,\Bigl({2p_\perp^2\over m_{H\perp}^2}\Bigr)
\biggl[\sqrt{b}W(b)-\sqrt{a}W(a)\biggr]
\cr&+\half\Bigl(1-{4m_t^2\over m_{H\perp}^2}\Bigr)
\biggl[W(b)^2-W(a)^2\biggr]
\Biggr\}\ ,}\eqn\ffs$$
where $a=1+4m_t^2/p_\perp^2$, $b=1-4m_t^2/m_H^2$, and we take the
root $\sqrt{b}=i\sqrt{|b|}$ for $b<0$.  We have also defined the
function
$$
W(c)\ =\ \cases{\phantom{\bigg[}\!
-2i\arcsin{(1/\sqrt{1-c})}\ ,&$\qquad\qquad
c<0$\cr\phantom{\bigg[}\!
\ln\displaystyle{1+\sqrt{c}\over1-\sqrt{c}}
\,-\,i\pi\ ,&$\qquad\qquad 0<c<1$\cr
\phantom{\bigg[}\!
\ln\displaystyle{\sqrt{c}+1\over\sqrt{c}-1}
\ ,&$\qquad\qquad c>1$\cr}\eqn\wi$$
Note that
${\cal F}_s(p_\perp^2)$ is proportional to the form factor $A_5$ of
Ref.~\EHSV\ in the appropriate kinematic channel.  In the limit
$p_\perp,m_H<<m_t$, the form factor goes to the value
${\cal F}_s=-(4/3)S$.

{\it ii)}\ \undertext{Pseudoscalar Higgs}.
The pseudoscalar Higgs coupling to the top quark is $-P(m_t/v)
A\bar\psi i\gamma_5\psi$.
In this case the form factor is
$$
{\cal F}_p(p_\perp^2)\ =\ P\,\Bigl({4m_t^2\over m_{H\perp}^2}\Bigr)
\bigl(-\half\bigr)
\biggl[W(b)^2-W(a)^2\biggr]
\ .\eqn\ffp$$
In the limit $p_\perp,m_H<<m_t$, we find ${\cal F}_p=2P$.

\endpage
\refout
\endpage
\figout
\endpage
\end